\documentstyle[psfig,epsf]{europhys}
\newcommand{\beq}{\begin{equation}}
\newcommand{\eeq}{\end{equation}}
\newcommand{\beqa}{\begin{eqnarray}}
\newcommand{\eeqa}{\end{eqnarray}}
\begin{document}

\euro{}{}{}{}
\Date{}
\shorttitle{A. PRIEL et.\ al.\ ROBUST CHAOS GENERATION BY A PERCEPTRON}
\title{Robust chaos generation by a perceptron}

\author{A. Priel\footnote{Present address: School of Mathematical Sciences,
        Tel-Aviv University, Tel-Aviv, Israel} and I. Kanter}
\institute{
        Minerva Center and Department of Physics, \\ 
        Bar-Ilan University, 52900 Ramat-Gan, Israel
}


\pacs{
\Pacs{84}{35+i}{Neural networks}
\Pacs{05}{45-a}{Nonlinear dynamics and nonlinear dynamical systems}
        }

\maketitle

\begin{abstract}
The properties of time series generated by a perceptron with
monotonic and non-monotonic transfer function, where
the next input vector is determined from past output values, are examined. 
Analysis of the parameter space reveals the following main finding:
a perceptron with a monotonic function can produce fragile chaos only whereas
a non-monotonic function can generate robust chaos as well. 
For non-monotonic functions, the dimension of the attractor can be controlled 
monotonically by tuning a natural parameter in the model.
\end{abstract}

Attractor neural networks composed of  
symmetric interactions were studied using methods closely 
related to those used in statistical physics  for the study of magnetic 
systems \cite{amit89}. Chaotic dynamics, however, can occur in asymmetric 
networks which are more relevant to the structure of biological systems.
For instance, analysis of experimental results obtained from EEG recordings 
revealed chaotic attractors \cite{babloyantz85}, however,
the role of this dynamic behavior in the brain is not yet clear. 
The emergence of chaos in some limited recurrent 
asymmetric networks was examined  either analytically or 
numerically \cite{asymnet}.

Another important class of models with an underlying dynamics 
which exhibits a chaotic behavior are time-delayed neural networks.
These networks are analogous to the non-linear auto-regressive (NAR) models
that are widely used for time series analysis. Some of their stationary 
statistical properties were reviewed by H.\ Tong \cite{tong90}. 
This letter is concerned with noiseless NAR networks where the non-linearity 
is realized through a feed-forward network. The computational capabilities of
such models were found to be equivalent to fully recurrent networks 
\cite{siegelman97}. A similar dynamical system, the generalized shift map, 
can be viewed as a special case of the noiseless NAR model in which piecewise
linear mappings are used. A proposed physical realization of the generalized 
shift map is demonstrated by the motion of a particle constrained 
by a system of parabolic mirrors in a 3D billiard \cite{moore90}.
 
In spite of the advances in the research of these models, the problem of 
learning a chaotic time series is still an open question and analytical 
results are hardly available. The interplay, if any, between the details
of the network ({\it i.e.}, its architecture, type of non-linearity and 
parameters) and the characteristics of the chaos generated by the model 
({\it e.g.}, structural stability, attractor dimension) is 
of both theoretical and practical interest. 
The simplest NAR network is clearly a perceptron, described
mathematically as follows. For a given input vector at time step $t$,
${\vec S}^t$ ($S_j^t , j=1,\ldots,N$), the network's output $S_{out}^t$ 
is given by
\beq\label{s_out}
S_{out}^t=f \left( \beta \sum_{j=1}^N  W_j  S_j^t \right)
\eeq
\noindent
where $\vec{W}$ are the weights, $\beta$ is a gain parameter and $f$ is a 
transfer function. The input vector at time $t+1$ is obtained by shifting 
the previous output values
\beq\label{dyn_rule}
S_1^{t+1} = S_{out}^t \quad ; \qquad S_j^{t+1}=S_{j-1}^t
\quad j=2, \ldots ,N \quad .
\eeq
\noindent
This model exhibits (quasi) periodic attractors in the stable regime,
regardless of the details of the weights \cite{sgen_list}. 
The same conclusion holds when the network is a multilayer-perceptron 
(altering eq.\ \ref{s_out} accordingly) \cite{sgen_mln}.

One of the main issues that we address in this letter is the interplay 
between the type of non-linearity and the stability of the chaos generated. 
In the sequel we present a numerical investigation of the NAR-perceptron 
using two types of transfer functions: monotonic and non-monotonic. 
Understanding the behavior of this simple model is essential before turning 
to the analysis of more complicated architectures. The distinction between
monotonic and non-monotonic functions was already pointed out in 
\cite{sgen_chaos_pre99,caroppo99}.
Another important question that arises is the capability to control the 
attractor dimension of the sequence generated by the model. The ability to
synthesize a chaotic neural network is clearly a challenging task.

In order to characterize the dynamical properties of the networks,
we classify the attractors in phase space in the vicinity of 
a given vector of parameters while varying some control parameters.
In order to minimize the number of free parameters in the model, 
we represent the weights in the Fourier domain and take only few components. 
This representation
is clearly advantageous in high dimensional systems where varying a large 
number of parameters is not practical and can not be visualized conveniently.
The general prescription of the weights is given by:
\beq\label{weights_general}
W_j = \sum_p a_p \cos({2\pi \over N} k_p j + \pi \phi_p) + b ~; \qquad
j=1 \ldots N , \quad \phi_p \in [-1..1] \quad ,
\eeq
\noindent
where $\{ a_p \}$ are constant amplitudes; $\{ k_p \}$ are positive integers 
denoting the wave numbers; $\{ \phi_p \}$ are the phases;
$b$ is the bias term and $p$ runs over the number 
of Fourier components composing the weights. 
We investigate only the cases $p=1$ and $2$, or randomly chosen weights.
The analysis of the first question is exemplified by the hyperbolic-tangent
function (monotonic) and the sine function (non-monotonic). 
We emphasize that other functions were tested as well, leading to the same 
conclusions ({\it e.g.}, monotonic and non-monotonic piecewise linear 
functions).

Starting with the monotonic functions, the output, $S_{out}$, is given by
\beq\label{tanh_output}
S_{out}^t=\tanh(\beta \sum_{j=1}^N W_j S_j^t ) \quad .
\eeq
\noindent
For weights that consist of a single biased Fourier component we set $a=1$.
In the simplest case of two inputs, $N=2$, 
the special case $~\phi=1~$ results in the following map
\beq\label{eq_sgen_N2}
S^{t+1}= \tanh \left[ \beta \left( (1+b) S^t + (-1+b) S^{t-1} \right) 
\right] \quad ,
\eeq
\noindent
which is equivalent to a physical model of a magnetic
system, the ANNNI model \cite{annni}.
This map is capable of generating stable attractors (fixed points, 
periodic and quasi-periodic orbits) as well as unstable chaotic 
(however, not robust) behavior.

\begin{figure}
\centerline{\psfig{figure=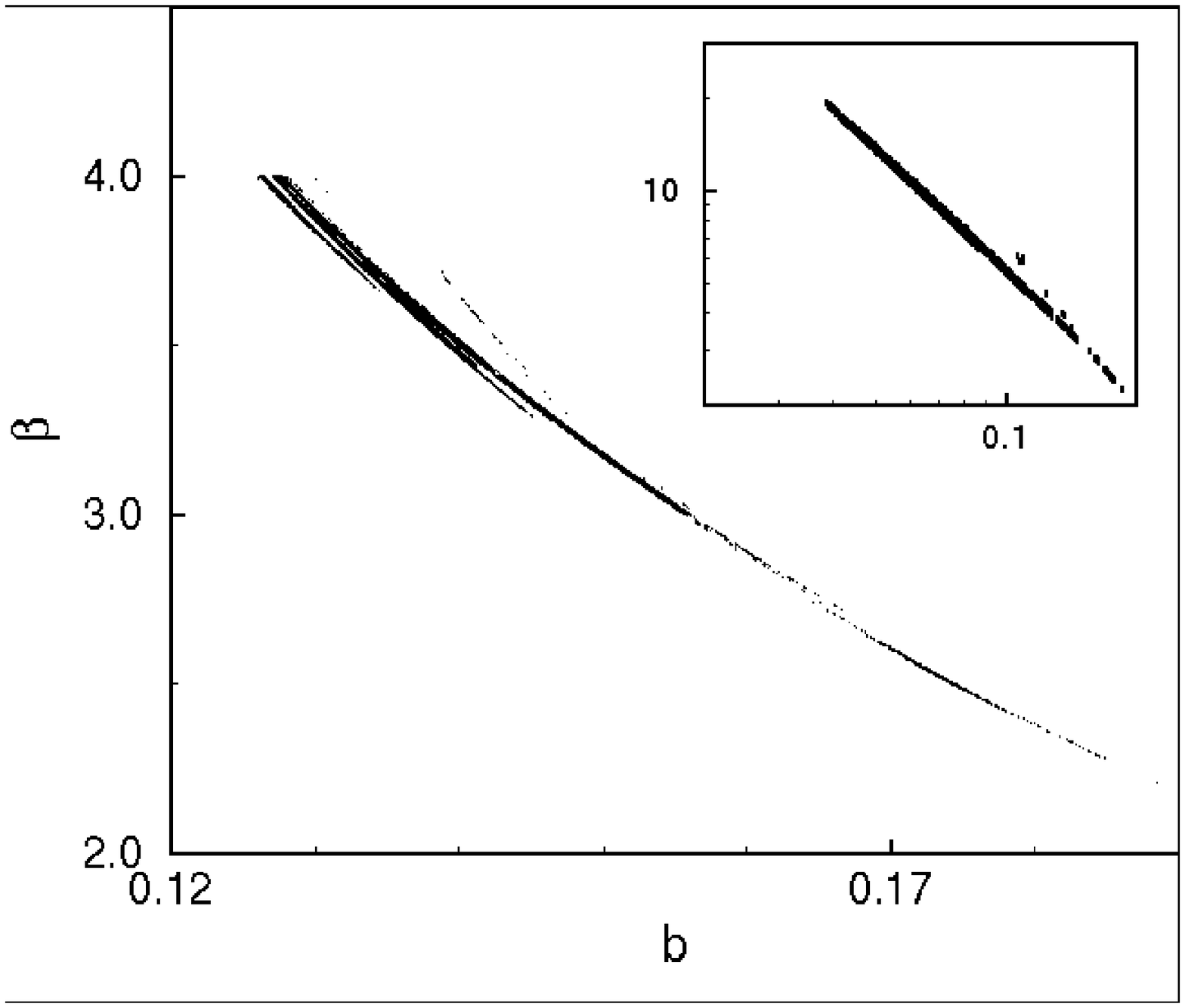,width=7.2cm}}
\caption{A region in parameter space for a NAR-perceptron with 
hyperbolic-tangent transfer function, $N=3$ and $\phi=1$ 
(eq.\ \ref{eq_sgen_N3}), where
vector points that lead to chaotic attractors are marked;
the remaining space in this region leads to stable attractors. Insert: 
a ($\log$-$\log$) continuation for higher gain values, indicating
a power-law relation $~\beta \propto b^{-1.32\pm 0.001}$.
}
\label{fig_th_N3}
\end{figure}

Using the same settings, the map for $N=3$ is given by:
\beq\label{eq_sgen_N3}
S^{t+1}= \tanh \left[ \beta \left( 
(1/2+b) S^t + (1/2+b) S^{t-1} + (-1+b) S^{t-2}\right) \right] 
\eeq
\noindent
which is similar to the ANNNI model with competing interactions 
between third neighbors along the axial direction \cite{moreira}.
In order to analyze numerically the parameter space of the map,
it is sampled in a high resolution (up to $10^{-5}$). 
The spectrum of Lyapunov exponents is estimated directly from the dynamic 
equations using an algorithm described 
in Wolf {\it et.\ al.\ } \cite{wolf_alg} and former authors 
(references therein).
Figure \ref{fig_th_N3} depicts a section of the parameter space of the map 
given in eq.\ \ref{eq_sgen_N3}. The black dots represent vector points that 
lead to chaotic behavior with one positive Lyapunov exponent; 
the remaining space in this region gives rise to stable attractors.
Each point has neighboring points which lead to periodic attractors. 
The insert suggests a possible power-law relation between the gain and the 
bias terms.
The general case $\phi<1$ can be analyzed in the same manner as that 
described above. We note that for $N=2$, no unstable regions are found 
for $\phi<{1 \over 2}$. 

\begin{figure}
\centerline{\psfig{figure=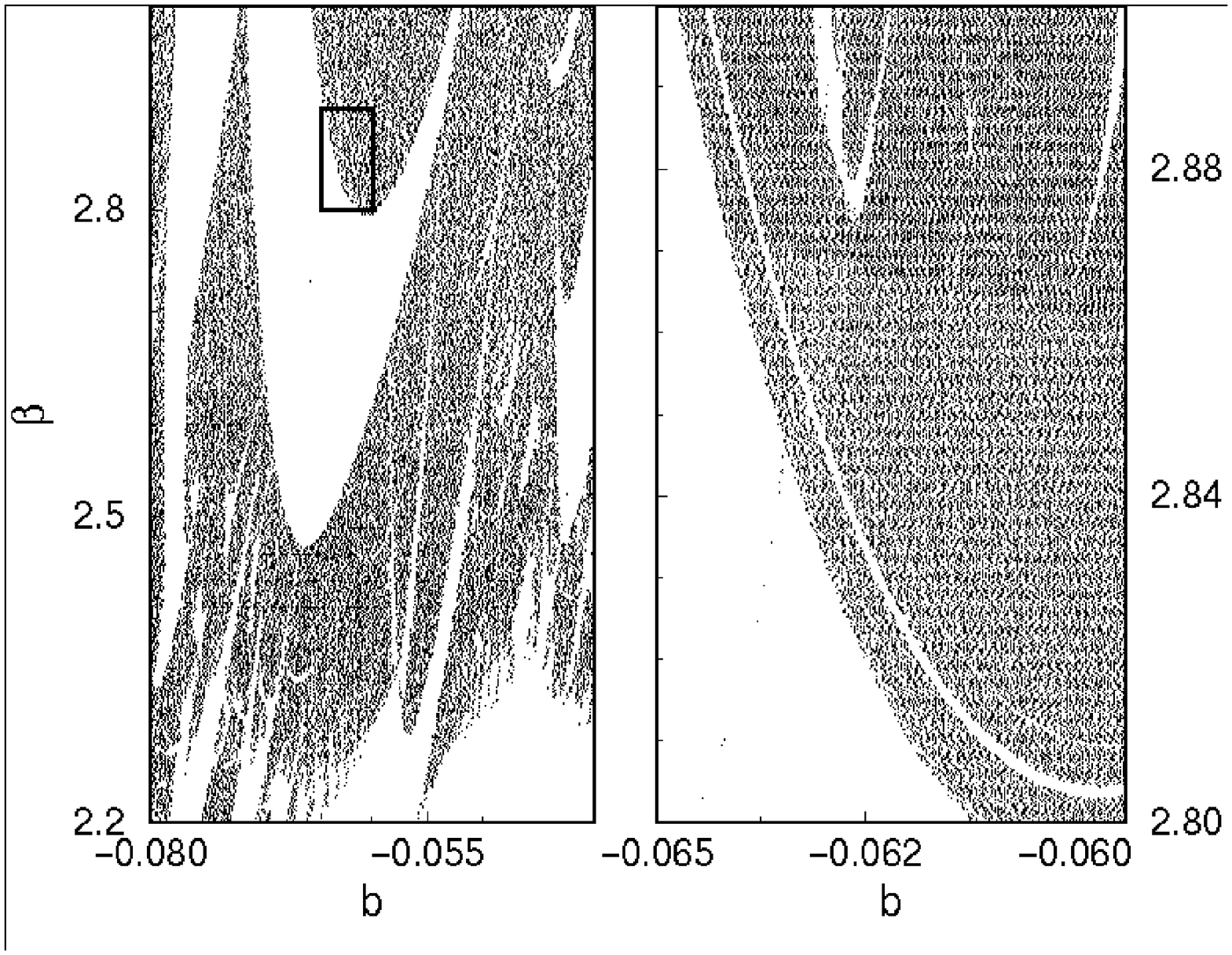,width=7.2cm}}
\caption{Example of a region in parameter space for hyperbolic-tangent 
transfer function and $N=9$, where vector points that lead to chaotic 
attractors are marked; the remaining space in this region leads to stable 
attractors. The right-hand part is a blowup of the marked rectangle.
}
\label{fig_th_N9}
\end{figure}

In higher dimensions (larger $N$), the parameter space becomes more 
structured. We demonstrate this behavior for $N=9$. 
As before, we restrict the dimension of parameter space to two, ($\beta,b$), 
with weights prescribed by:
\beq\label{weights_N9}
W_j = \cos({2\pi \over N} j + \pi \phi) + b ~; \qquad
j=1 \ldots N , \quad \phi=1, \quad N=9 \quad .
\eeq
\noindent
Figure \ref{fig_th_N9} depicts a region in parameter space for the case 
$N=9$ with $\phi=1$ (unstable behavior is found outside this region as well).
We use $\phi=1$ here as well since larger phase generates more
unstable points. The unstable points span a significant part of the space.  
Qualitatively, the parameter space is similar to that of $N=2,3$ in the sense 
that the chaotic attractors are fragile with a single positive exponent. 
However, as $N$ increases, the structure of the parameter space
becomes more involved as larger cycles become available.
Moreira and Salinas \cite{moreira} have already mentioned that such a 
complication is expected at larger $\beta$ in their model ($N=3$). 
We should stress here that the regions of unstable points are not dense;
{\it i.e.}, in the vicinity of every unstable point there exists a stable one,
therefore, the chaotic regions are fragile. 
In particular, these points generate a mixed behavior in phase 
space; stable and unstable attractors are possible, depending on the initial 
condition, and both have a non-vanishing basin of attraction.
The examples provided so far consist of weights with a single 
dominant Fourier component in the power spectrum.
We tested the case of two Fourier components with
bias of the form:
\beq\label{weights_2p}
W_j = a_1 \cos\left({2\pi \over N} k_1 j + \pi \phi_1\right) + 
a_2 \cos\left({2\pi \over N} k_2 j + \pi \phi_2 \right) + b 
\quad .
\eeq
\noindent
Many cases with arbitrary amplitudes ($a_1,a_2$) and phases ($\phi_1,\phi_2$) 
were examined exhaustively; the wave numbers ($k_1,k_2$) were kept constant 
for each $N$ tested. The results indicate 
that our conclusions are applicable in the more general case. 
In all our simulations, we found no regions
with more than a single positive exponent (for $N$ up to $60$), including
many cases with randomly chosen weights ({\it i.e.} arbitrary number of
Fourier components).

Based on these results, we conjecture that the perceptron with a monotonic 
transfer function typically exhibits unstable behavior with a single positive 
Lyapunov exponent. Note that the bias term $b$, is
crucial for producing chaotic behavior with monotonic 
transfer functions. Another important ingredient is the existence of a large
enough phase, at least when the weights consist of a single Fourier 
component. It is possible that
additional Fourier components are sufficient to generate unstable 
behavior (without large phase), however, larger phase 
significantly increases the number of unstable points. \\

Applying a non-monotonic transfer function gives rise to a different 
repertoire of properties in the parameter space 
(with respect to monotonic functions). 
One observes {\it robust chaos}, and the number
of positive exponents can be larger than one. 
In the following analysis we use the sine as a representative function. 
At low gain values, quasi-periodic stationary 
solutions were found analytically, see \cite{sgen_mln};
in the following we focus on the region of high gain values where unstable
behavior emerges. Note that in contrast to monotonic functions, 
unstable dynamics can be obtained with phase and bias equal to zero. 
Indeed, in the sequel we use $\phi=0$; the parameter space remains two 
dimensional, $\beta$-$b$, as above.  

\begin{figure}
\centerline{\psfig{figure=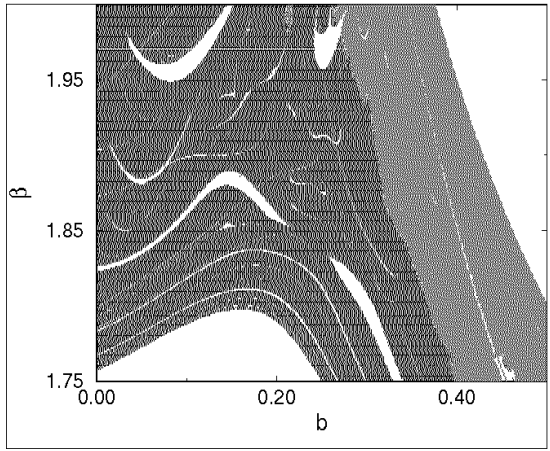,width=7.2cm  }}
\caption{Analysis of a region in parameter space for a network with
a sine transfer function and $N=2$, where points that lead to chaotic 
trajectories are marked. 
The dark(gray) colors correspond to areas with one(two) positive exponent;
the remaining space in this region leads asymptotically to stable attractors. 
}
\label{fig_sin_N2_general}
\end{figure}

\begin{figure}
\centerline{\psfig{figure=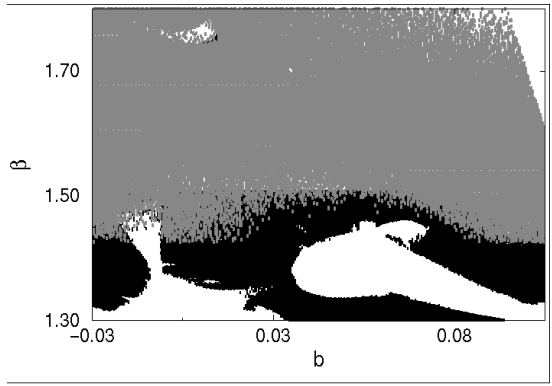,width=7.2cm  }}
\caption{Analysis of a region in parameter space for a network with
a sine transfer function and $N=9$.
The dark area correspond to one or two positive exponents; the gray area
corresponds to three positive exponents.
The remaining space in this region leads asymptotically to stable attractors. 
}
\label{fig_sin_N9}
\end{figure}

Let us first describe the numerical analysis. A region in parameter space,
for which the spectrum of Lyapunov exponents has been calculated, was sampled 
exhaustively in a resolution of $ \approx 10^{-5} ~$.
Several random initial conditions were used for each vector in the parameter 
space to avoid possible isolated limit cycles. 
Figure \ref{fig_sin_N2_general} depicts such
a region for $N=2$, where vector points that lead to chaos are marked;
vectors in white areas lead asymptotically to stable attractors.
The dark area corresponds to a region with
one positive exponent, while the gray area corresponds to a region with
two positive exponents. 
A perusal in the figure reveals the following characteristics in the 
structure of the parameter space. The dark region (l.h.s.\ of the figure) 
contains extended (stable) windows (embedded white areas)
associated with cycles of different length.
The common feature of these windows is the fact that they are surrounded
by unstable regions with one positive exponent. 
As we move to the right-hand side of the figure, a region with two positive 
exponents emerges; however, this type of trajectories is less interesting 
since the volume in phase space is expanding, hence, the bounded space is 
filled.
The case $N>2$ reveals another aspect in the structure of parameter space.
There are regions for which we find more than two positive Lyapunov exponents.
In such regions, we observed a robust chaos, namely, small changes of 
the parameters would not destroy the chaotic behavior. 
Figure \ref{fig_sin_N9} depicts the parameter space for $N=9$ with weights
prescribed by eq.\ \ref{weights_N9} and $\phi=0$. The dark area corresponds 
to a region with one or two positive exponents; the gray area corresponds
to three positive exponents. In the gray area we observed a
robust chaos with volume contracting attractors.

Let us now extend our analysis for large systems. Two questions come to the
fore: \\ (1) Can we find regions in which the chaotic dynamics is
robust, and how frequent are they~? \\
(2) How does the attractor dimension relate to the control parameters~?

\begin{figure}
\centerline{\psfig{figure=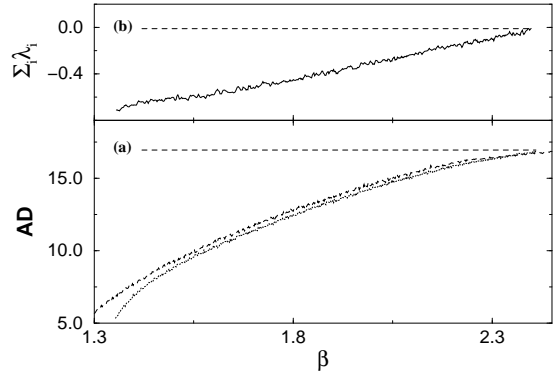,width=7.2cm  }}
\caption{(a) The attractor dimension (AD) as a function of the gain
for a network with sine transfer function, $N=17$, $\phi=0$,
$b=0$ -  dashed line and $b=-0.05$ - long dashed line.
The horizontal dashed line is at ~AD=17.
(b) The sum of Lyapunov exponents for $b=0$.
The horizontal dashed line is at ~$\sum_{i=1}^{N} \lambda_i=0$
}
\label{fig_ad_sumexp_N17_sin}
\end{figure}

We claim that the possibility of discovering
regions with an increasing number of positive exponents, grows with $\beta$.
This means that we have a natural parameter in the model that controls the 
structure of the chaos and the attractor dimension.
In order to test this conjecture, we used a larger system, $N=17$. 
For this analysis we used more complicated weights, consist of two 
Fourier components with irrational phases and a bias term 
(eq.\ \ref{weights_2p}). 
The amplitudes and phases of the components were kept fixed, 
therefore, we have the same two dimensional parameter space, as before. 
The qualitative results reported below are insensitive to the exact values 
of the amplitudes and phases.
A close inspection of the parameter space reveals the following regimes:
First, the incommensurate regime which corresponds to the irrational 
phases of the weights. 
Above some value of the gain parameter (depending on the details of
the weights) most of the space is associated with chaotic dynamics and the 
number of positive exponents grows with $\beta$ (when the sum of the 
exponents becomes positive the dynamics is area expanding). 
In this regime, we observed a 
relatively monotonic growth of the attractor dimension. The estimation
of the attractor dimension was done using the Kaplan-Yorke conjecture 
\cite{kap_yorke}. Figure \ref{fig_ad_sumexp_N17_sin}(a) depicts the attractor 
dimension as a function of $\beta$ for two fixed bias values. 
Clearly, the dimension grows monotonically. 
Figure \ref{fig_ad_sumexp_N17_sin}(b) shows the sum of the exponents, 
~$\sum_{i=1}^{N} \lambda_i$~, in the same region of gain values.
The sum grows monotonically, as expected, and saturates zero where the 
attractor dimension saturates the dimension of the system, $N$.
Each point was averaged over $10$
random initial conditions in order to check whether the same attractor is 
sampled. Indeed, the errors are less than $1 \%$ and typically much less,
therefore, they are not presented.
Note that there are cases, not shown in this figure, for which the line 
$b=\mbox{const}$ crosses a (stable) window; in these cases, the attractor 
dimension decreases and then continues to grow once the window is passed. 
Finally, we validated these results by estimating the attractor 
dimension from the time series generated by the network, and compared 
it to the estimation using the Kaplan-Yorke conjecture.
The time series was recorded at the same time the spectrum of exponents was 
evaluated; the attractor dimension was calculated from the
reconstructed phase space using the method of Correlation-Integral 
\cite{grassberger}. The results confirm our conjecture for the monotonic 
relation between the attractor dimension and $\beta$. 

The major conclusion of this letter is that a NAR-perceptron model with 
non-monotonic transfer function can generate a stable chaotic attractor. 
The attractor dimension can be controlled by varying a parameter in the model.
At this stage, we can not rule out the possibility that extremely tiny stable
windows are situated in the vicinity of every chaotic point, however, this 
is very unlikely to be the case when the number of positive exponents is much
larger than the number of free parameters. In fact the larger $N$ is, the less
probable are the stable windows in these regions (see also \cite{barreto}).
Further analytical study of piecewise linear functions may clarify this issue. 
It is interesting to mention that Banerjee 
et.\ al.\ \cite{banerjee} showed a robust chaos in a simple 2D non-monotonic 
piecewise linear map. 

Training a model to generate robust chaos seems to be limited to 
non-monotonic transfer functions, and may be of practical interest  
where reliable operation is necessary.


\end{document}